# EQUILIBRIUM CONSTANTS OF NUCLEAR REACTIONS IN SUPERNOVA EXPLOSIONS

*Jorge A. Muñoz, Marcos A. García and Jorge A. López[*]*

Department of Physics, University of Texas at El Paso, El Paso, TX 79968-0515, U.S.A.

**Abstract**

We study the change in internal rotational energy in the transformation of protons to neutrons in the β-decay reactions that take place in the collapse of the iron core of massive stars that precede type II supernova explosions. We consider an ensemble of electrons, protons, neutrons and neutrinos undergoing β-decay reactions, treat the protons and neutrons as triatomic rotors, evaluate the equilibrium constant to obtain the change in rotational energy during the proton-to-neutron transformation. We estimate such change for a variety of conditions, and compare to the energy released in a supernova explosion

**Resumen**

Estudiamos el cambio en la energía de rotación interna en la transformación de protones a neutrones en las reacciones de decaimiento β que tienen lugar en el colapso del núcleo de hierro de estrellas masivas que precede a explosiones de supernova tipo II. Consideramos un ensamble de electrones, protones, neutrones y neutrinos sometidos a reacciones de decaimiento β, tratamos los protones y neutrones como rotores triatomicos, evaluamos la constante de equilibrio, y obtenemos el cambio en la energía rotacional durante la transformación de protones a neutrones. Calculamos este cambio para una variedad de condiciones, y comparamos con la energía liberada en una explosión de supernova

**Palabras clave:** función de partición rotacional, constante de equilibrio, supernova tipo II, decaimiento β, energía libre de Gibbs

## I. INTRODUCTION

Type II Supernovae result from the collapse of the star's iron core. For stars with iron core masses smaller than the Chandrasekhar limit of 1.4 solar masses, the electron degeneracy pressure is enough to prevent the further gravitational collapse, but for larger core masses, protons and electrons in the core (in iron atoms) are forced together to form neutrons and neutrinos. As this happens, the core collapses and the outer layers of the star fall inward due to gravity. This continues filling the core with neutrons (the neutrinos escape as they don't interact strongly with matter), until a neutron degenerate state is formed, which produces enough pressure to stop the collapse. At that point the falling outer layers crash into the core and "bounce" outwards, creating a shock wave. The outward pressure from the neutrinos and the shock wave cause the outer layers of the star to blow itself apart in a Type II supernova, leaving the core as a remnant known as a neutron star.

The nuclear reaction responsible for the collapse is the $\beta$-decay reaction:

[*]Corresponding Author: jorgelopez@utep.edu

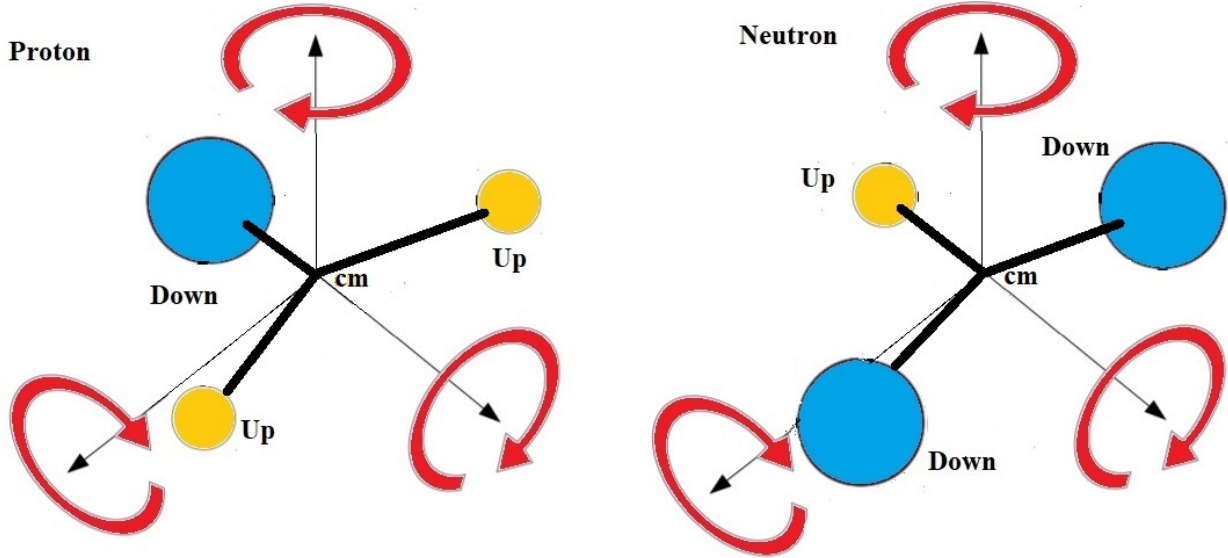

Figure 1. Model tri-quark rigid rotor composition of a proton and a neutron. "cm" stands for center of mass.

$$e^- + p \rightarrow n + v_e,$$

where $e^-$, $p$, $n$ and $v_e$ are, respectively, the electron, proton, neutron and electron neutrino. It is known, since the work of Bethe and collaborators [1,2], that the electron capture does not occur with free protons, but with protons in $^{56}$Fe nuclei by means of Gamow–Teller transitions which dominate electron capture at low excitation energies. In fact, Ref. [2] studied the transformation of an f7/2 proton in $^{56}$Fe into an f5/2 neutron by electron capture, which leaves the daughter nuclei excited and ready to decay to the ground state by γ -emission.

Here we are interested in studying the transformation of protons to neutrons using a simple classical parton model [3], in particular its change in internal rotational energy. Considering the protons and neutron as triatomic rigid "molecules" each containing 3 quarks (see Figure 1), one can see that since their internal structures are different, their rotational energies must differ. How large is the change in rotational energy?

For our calculation, we consider a large system of electrons, protons, neutrons and neutrinos undergoing a reversible $\beta$-decay reaction

$$e^- + p \rightleftharpoons n + v_e,$$

We then use the law of mass-action of such reaction to obtain the equilibrium constant, and finally use the equilibrium constant to obtain the change in Gibb's free energy during the proton-to-neutron massive transformation. The equilibrium constant will help us understand whether supernova explosions will have a higher concentration of either protons and electrons or neutrons and electron neutrinos at the supernova's point of equilibrium. In the following section we implement these calculations.

## II. METHODOLOGY

We consider the protons and neutron to be triatomic rigid "molecules" each containing 3 quarks with effective masses. The total number of degrees of freedom is 3×3, with 3 for the translational motion of the nucleon as a whole, 3 rotational degrees of freedom, and 3×3 − 6 = 3 vibrational degrees. We focus on the rotational degrees of freedom. The up and down quarks have different masses, so by their symmetry both protons and neutrons can be considered as symmetric tops. The effective masses phenomenologically account for the quantum chromodynamics binding energy (energy of the gluon fields and kinetic energy) responsible for most of the hadron mass. We get the effective masses of the up quark $M_u$ and of the down quark $M_d$ by solving the following system of equations

$$M_u + 2M_d = 939.56 \text{ MeV}/c^2$$
$$2M_u + M_d = 938.27 \text{ MeV}/c^2.$$

The equations correspond to the components and mass of the neutron and the proton, respectively. The result is $M_u = 312.33 \text{ MeV}/c^2$ and $M_d = 313.62 \text{ MeV}/c^2$. Notice that the effective masses are much greater than the rest masses of the up and down quarks since the rest masses of quarks are a very small fraction of the hadron mass.

From statistical mechanics, and under the assumption that the spectrum of the rotational energy of these particles is dense, their rotational partition functions can be calculated by

$$Z(T) = \sum_i g_i e^{-\frac{E_i}{K_B T}},$$

where the sum is over all rotational energy states, $E_i$, each of which has a degeneracy $g_i$. As explained in the Appendix, for a symmetric top, two of the three moments of inertia are equal, and the classical rotational partition function is given by:

$$q_{rot} = \frac{\pi^{1/2}}{\sigma} \left(\frac{8\pi^2 k_B T I_B}{h^2}\right) \sqrt{\frac{8\pi^2 k_B T I_A}{h^2}},$$

where $T$ is the temperature of the system, $k_B$ is the Boltzmann constant, $h$ is the Planck constant, $I_A$ and $I_B$ are the principal moments of inertia, and $\sigma$ is the symmetry number that corrects for overcounting rotational configurations.

The equilibrium constant of the nuclear reaction is the value of the reaction quotient at equilibrium, and it can be calculated through the law of mass-action:

$$\pi_j n_j^{\nu_j} = K(\tau) = \pi_j n_{Q_j}^{\nu_j} nq_{int}^{\nu_j},$$

which for the $\beta$-decay reaction yields

$$\frac{1}{K^{\ominus}} = \frac{\left(\frac{M_p \tau}{2\pi \hbar^2}\right)^{\frac{3}{2}} \left(\frac{M_e \tau}{2\pi \hbar^2}\right)^{\frac{3}{2}} q_p}{\left(\frac{M_n \tau}{2\pi \hbar^2}\right)^{\frac{3}{2}} \left(\frac{M_{\nu_e} \tau}{2\pi \hbar^2}\right)^{\frac{3}{2}} q_n} = \frac{(M_p)^{\frac{3}{2}} (M_e)^{\frac{3}{2}} q_p}{(M_n)^{\frac{3}{2}} (M_{\nu_e})^{\frac{3}{2}} q_n},$$

where $q_n$, and $q_p$, are the rotational partition functions (sum over states) of the neutron and proton, respectively. Of course, for the electron and neutrino, being point particles, the partition functions $q_{v_e}$ and $q_{e^-}$ are equal to 1, while for the protons and neutrons it must be calculated by summing over their rotational energies. Thus,

$$K^\ominus = \frac{q_n q_{v_e}}{q_p q_{e^-}} = \frac{q_n}{q_p} = \frac{\frac{\pi^{1/2}}{\sigma}\left(\frac{8\pi^2 k_B T I_B^{(n)}}{h^2}\right)\sqrt{\frac{8\pi^2 k_B T I_A^{(n)}}{h^2}}}{\frac{\pi^{1/2}}{\sigma}\left(\frac{8\pi^2 k_B T I_B^{(p)}}{h^2}\right)\sqrt{\frac{8\pi^2 k_B T I_B^{(p)}}{h^2}}} = \frac{I_B^{(n)}\sqrt{I_A^{(n)}}}{I_B^{(p)}\sqrt{I_A^{(p)}}} \; .$$

To calculate the moments of inertia we used the Atomic Simulation Environment (ASE) code [4] with the effective masses calculated above. Since the radii of both the proton and the neutron are not exactly known, we used combinations of radii from $0.8\times10^{-15}$ m to $1.15\times10^{-15}$ m with the increments of $1\times10^{-17}$ m, which covers the whole possibility space (see, e.g., Refs. [5-7]).

## III. RESULTS AND DISCUSSION

Evaluating the previous expression for the nuclear case yields

$$\frac{1}{K^\ominus} = 1.1 \times 10^8 \frac{q_p}{q_n}.$$

Table 1 shows the values of $K^\ominus$ for different ratios of the rotational partition functions of protons and neutrons, and at varying proton radius. It is estimated that the energy released in a core collapse supernova explosion is about $1\times10^{46}$ J [8, 9], so we can assess our model by calculating the change in the Gibbs free energy $\Delta G^\ominus$ when the reaction occurs at temperature $T$, given by

$$\Delta G^\ominus = -RT \ln K^\ominus ,$$

where $R$ is the universal gas constant. The temperature during a core collapse supernova explosion is about 100 billion kelvin [10] and our values for $K^\ominus$ range between $2.60\times10^{-8}$ and $0.89\times10^{-8}$, so the $\Delta G^\ominus$ predicted by our model is between $1.45\times10^{13}$ J/mol and $1.54\times10^{13}$ J/mol. If the number of particles that participate in the reaction is similar to the number of particles in the Sun (about $1.2\times10^{57}$ particles), then about $2\times10^{33}$ moles of particles will participate and the $\Delta G^\ominus$ predicted by our model is between $2.9\times10^{46}$ J and $3.1\times10^{46}$ J.

Evidently, the model that we present here is very simple, classical, and hides away much of the physics. Nevertheless, the predicted energy for a core collapse supernova explosion is in the correct order of magnitude and only about 3 times as large as that estimated by the most sophisticated currently available methods. Several improvements could be made to the model, for example, it is known that both protons and neutrons have up and down quark distributions ("shapes") [10] that are dependent on the hadron energy [11, 12], so the moment of inertia could be calculated with a distribution of a large of number of partons as opposed to only 3 representing the up and down quarks. Here we considered rigid rotors, but it is likely that vibrational degrees of freedom play a role as well. Perhaps the most interesting result is the suggestion that the "shape" of the protons and neutrons might play a role in the $\beta$-decay reaction at high energies.

*Table 1*. Values of $K^{\ominus}$ for different ratios of protons to neutrons, and at radii varying between $0.8 \times 10^{-15}$ m to $1.15 \times 10^{-15}$ m in steps of about $5 \times 10^{-17}$ m.

| $q_p/q_n$ | Proton radius (m) | $K^{\ominus}$ |
|---|---|---|
| 2.88 | $0.80 \times 10^{-15}$ | $2.60 \times 10^{-8}$ |
| 2.49 | $0.84 \times 10^{-15}$ | $2.25 \times 10^{-8}$ |
| 2.09 | $0.89 \times 10^{-15}$ | $1.89 \times 10^{-8}$ |
| 1.78 | $0.94 \times 10^{-15}$ | $1.60 \times 10^{-8}$ |
| 1.52 | $0.99 \times 10^{-15}$ | $1.37 \times 10^{-8}$ |
| 1.31 | $1.04 \times 10^{-15}$ | $1.18 \times 10^{-8}$ |
| 1.14 | $1.09 \times 10^{-15}$ | $1.03 \times 10^{-8}$ |
| 0.99 | $1.14 \times 10^{-15}$ | $0.89 \times 10^{-8}$ |

## IV.  CONCLUSION

We studied the equilibrium constant of the $\beta$-decay reaction in core collapse supernova explosions by assuming that protons and neutrons are rigid-rotor symmetric tops with a classical rotational partition function determined by classical moments of inertia. Using this simple classical model we obtain a prediction for the energy of a core collapse supernova explosion that is in the correct order of magnitude comparable to more advanced models. Our results can be improved by considering more refined internal proton and neutron structures, and by incorporating vibrational degrees of freedom as well. Our work raises the question if the internal "shape" of protons and neutrons have a say in the β-decay reactions that precede the Type II supernova explosions.

## V.  REFERENCES

1. H.A. Bethe, Rev. Mod. Phys. 62 (1990) 801.

2. H.A. Bethe, G.E. Brown, J. Applegate, J.M. Lattimer, Nucl. Phys. A 324 (1979) 487.

3. R. P. Feynman. In: M. E. Noz, Y. S. Kim (eds) Special Relativity and Quantum Theory. Fundamental Theories of Physics (An International Book Series on The Fundamental Theories of Physics: Their Clarification, Development and Application), vol 33. Springer, Dordrecht.

4. Ask Hjorth Larsen, Jens Jørgen Mortensen, Jakob Blomqvist, Ivano E. Castelli, Rune Christensen, Marcin Dułak, Jesper Friis, Michael N. Groves, Bjørk Hammer, Cory Hargus, Eric D. Hermes, Paul C. Jennings, Peter Bjerre Jensen, James Kermode, John R. Kitchin, Esben Leonhard Kolsbjerg, Joseph Kubal, Kristen Kaasbjerg, Steen Lysgaard, Jón Bergmann Maronsson, Tristan Maxson, Thomas Olsen, Lars Pastewka, Andrew Peterson, Carsten Rostgaard, Jakob Schiøtz, Ole Schütt, Mikkel Strange, Kristian S. Thygesen, Tejs Vegge,

Lasse Vilhelmsen, Michael Walter, Zhenhua Zeng, Karsten Wedel Jacobsen. "The Atomic Simulation Environment—A Python library for working with atoms." J. Phys.: Condens. Matter. 29, 273002 (2017)

5. Pohl, R., Antognini, A., Nez, F. et al. The size of the proton. Nature 466, 213–216 (2010).

6. Beyer, Axel and Maisenbacher, Lothar and Matveev, Arthur and Pohl, Randolf and Khabarova, Ksenia and Grinin, Alexey and Lamour, Tobias and Yost, Dylan C. and Hansch, Theodor W. and Kolachevsky, Nikolai and Udem, Thomas. "The Rydberg constant and proton size from atomic hydrogen." Science 358, 79 (2017).

7. Filin, A. A. and Baru, V. and Epelbaum, E. and Krebs, H. and M\"oller, D. and Reinert, P. "Extraction of the Neutron Charge Radius from a Precision Calculation of the Deuteron Structure Radius." Phys. Rev. Lett. 124, 082501 (2020).

8. Janka, Hans-Thomas. "Explosion Mechanisms of Core-Collapse Supernovae." Annual Review of Nuclear and Particle Science 62, 407 (2012).

9. Smartt, Stephen J. "Progenitors of Core-Collapse Supernovae." Annual Review of Astronomy and Astrophysics 47, 63 (2009).

10. H.-Th. Janka, K. Langanke, A. Marek, G. Martínez-Pinedo, B. Müller. "Theory of core-collapse supernovae." Physics Reports 442, 38 (2007).

11. G. A. Miller. "Charge Densities of the Neutron and Proton." Phys. Rev. Lett. 99, 112001 (2007).

12. L. A. Harland-Lang, A. D. Martin, P. Motylinski, and R. S. Thorne. "Parton distributions in the LHC era: MMHT 2014 PDFs." Eur. Phys. J. C 75, 204 (2015).

13. G. A. Miller. "Shapes of the proton." Phys. Rev. C 68, 022201(R) (2003).

14. Larsen, D. (2016), *Physical Chemistry II: Properties of atoms and molecules*, Davis, CA, USA: LibreTexts. https://batch.libretexts.org/print/Letter/Finished/chem-2737/Full.pdf.

15. Jeschen, G. (2020), *Statistical thermodynamics*, Davis, CA, USA: LibreTexts. https://batch.libretexts.org/print/Letter/Finished/chem-2737/Full.pdf.

**APPENDIX: Rotational partition function of a triatomic molecule**

Protons and neutron can be considered triatomic "molecules" as each contains 3 quarks (see Figure 1). The number of degrees of freedom is 3×3, with 3 being translational of the nucleon as a whole, 3 rotational (shown in Figure 1), and the 3×3 – 6 = 3 vibrational. The rotational partition function of such a body is obtained in terms of the rotational energy levels, which in turn are obtained from the angular momenta, angular speed, and moments of inertial of the molecule. The formulation is simplified when it is calculated in the frame of reference of the principal axes of the nucleon in its center of mass.

The center of mass is defined by the equalities

$$\sum_{i=1,2,3} m_i x_i = \sum_{i=1,2,3} m_i y_i = \sum_{i=1,2,3} m_i z_i = 0,$$

where $m_i, x_i, y, z_i$ are the masses and coordinates of the three quarks in an arbitrary set of axes. The principal axes are obtained through a diagonalization of the moment of inertia tensor

$$\mathbf{I} = \begin{bmatrix} I_{xx} & I_{xy} & I_{xz} \\ I_{yx} & I_{yy} & I_{yz} \\ I_{zx} & I_{zy} & I_{zz} \end{bmatrix},$$

where the inertia products are $I_{xx} = \sum_i m_i (y_i^2 + z_i^2)$ and $I_{xy} = \sum_i m_i x_i y_i$, with similar expression for the other components. After the diagonalization, effectively a rotation of axes, all the off-diagonal terms will be zero and the resulting axes will be the principal axes with diagonal terms $I_A$, $I_B$, and $I_C$ as the principal moments of inertia. In that new set of Cartesian coordinates $\{x,y,z\}$, which pass through the center of mass of the body, the principal moments of inertia are:

$$I_A = \sum_{i=1}^{3} m_i [(y_i - y_{cm})^2 + (z_i - z_{cm})^2],$$

$$I_B = \sum_{i=1}^{3} m_i [(x_i - x_{cm})^2 + (z_i - z_{cm})^2],$$

$$I_C = \sum_{i=1}^{3} m_i [(x_i - x_{cm})^2 + (y_i - y_{cm})^2],$$

where the subscript "cm" stands for center of mass. It is clear that for the case depicted in Figure 1, i.e. for a rigid triatomic rotor, two of the principal moments of inertial will be equal to each other independent of how the axes are chosen.

Since the rotational kinetic energy of a diatomic rotor is $½I\omega^2 = L^2/2I$, where $\omega$ and $L$ are the angular speed and the angular momentum, the energy for a triatomic nucleon is

$$\varepsilon = ½I_A\omega_A^2 + ½I_B\omega_B^2 + ½I_C\omega_C^2 = \frac{L_A^2}{2I_A} + \frac{L_B^2}{2I_B} + \frac{L_C^2}{2I_C}$$

where $\omega_i$ and $L_i$ are the angular speeds and the angular momenta.

For a **symmetric top** two components of the moments of inertia are equal, $I_B = I_C$, and the rotational energy levels are specified by the angular momentum quantum number $J$, and the magnetic moment $K$, where $J$ takes on values 0, 1, 2, ... and $K = -J, -J+1, ... 0, 1, 2, ... J$.

The energy levels (with degeneracy $2J+1$) are given by

$$\varepsilon_{J,K} = \bar{B} J(J+1) + (\bar{A} - \bar{B}) K^2,$$

where $\bar{A} = h^2/8\pi^2 I_A$, and $\bar{B} = h^2/8\pi^2 I_B$, both in joules, and $h$ is Planck's constant. The partition function is (see e.g. Refs. [14, 15])

$$q_{rot} = \frac{1}{\sigma}\sum (2J+1)e^{-\varepsilon_{J,K}/k_BT} = \frac{1}{\sigma}\sum_{J=0}^{\infty}(2J+1)e^{-\bar{B}J(J+1)/k_BT}\sum_{K=-J}^{J}e^{-(\bar{A}-\bar{B})K^2/k_BT}$$

Where $k_B$ is the Boltzmann constant, $T$ is the temperature, and $\sigma$ is a factor that corrects for overcounting rotational configurations that emerge repeatedly due to symmetry: $\sigma = 1$ if the molecule does not have a center of symmetry, and $\sigma = 2$ if it does. If the rotational levels are lying very close to one another, the previous expression can be approximated by

$$q_{rot} = \frac{1}{\sigma}\int_0^{\infty}(2J+1)e^{-\bar{B}J(J+1)/k_BT}dJ\int_0^{\infty}e^{-(\bar{A}-\bar{B})K^2/k_BT}dK,$$

The first integral in the previous expression can be evaluated with the change of variables $x = J(J+1)$ and $dx = (2J+1)dJ$, and the second integral with $\int_{-\infty}^{\infty}e^{-ax^2}dx = \sqrt{\pi/a}$, to yield

$$q_{rot} = \frac{\pi^{1/2}}{\sigma}\left(\frac{8\pi^2 k_B T I_B}{h^2}\right)\sqrt{\frac{8\pi^2 k_B T I_A}{h^2}}.$$

The classical rotational partition function of an **asymmetric top** is more complex as all three moments of inertia and angular momenta must be taken into account. The partition function can be obtained by writing the rotational energy in terms of the Euler angles: $0 < \theta < 2\pi$, $0 < \varphi < 2\pi$, and $0 < \psi < 2\pi$:

$$q_{rot} = \int_{-\infty}^{\infty}\int_{-\infty}^{\infty}\int_{-\infty}^{\infty}\int_0^{\pi}\int_0^{2\pi}\int_0^{2\pi}\frac{1}{h^3}e^{-H(p,q)/k_BT}dp_{\theta}dp_{\phi}dp_{\psi}d\theta d\phi d\psi,$$

where $H(p,q)$ is the rotational Hamiltonian for kinetic energy, $k_B$ is the Boltzmann constant, $T$ is the temperature, $h$ and is Planck's constant. The rotational Hamiltonian for the kinetic energy can be written in terms of the Euler angles and their conjugate momenta, $p_{\theta}, p_{\phi}, p_{\psi}$ as follows

$$H(p,q) = \frac{\sin^2\psi}{2I_A}\left(p_{\theta} - \frac{\cos\psi}{\sin\theta\sin\psi}(p_{\phi}-\cos\theta p_{\psi})\right)^2 + \frac{\cos^2\psi}{2I_B}\left(p_{\theta} + \frac{\sin\psi}{\cos\theta\cos\psi}(p_{\phi}-\cos\theta p_{\psi})\right)^2 + \frac{1}{2I_C}p_{\psi}^2$$

This expression can be simplified by writing

$$\frac{H(p,q)}{k_BT} = \frac{1}{2I_Ak_BT}\left(\frac{\sin^2\psi}{I_A}+\frac{\cos^2\psi}{I_B}\right)\left(p_{\phi}+\left(\frac{1}{I_B}-\frac{1}{I_A}\right)\frac{\sin\psi\cos\psi}{\sin\theta\left(\frac{\sin^2\psi}{I_A}+\frac{\cos^2\psi}{I_B}\right)}(p_{\phi}-\cos\theta p_{\psi})\right)^2$$

$$+ \frac{1}{2k_BTI_AI_B\sin^2\theta}\left(\frac{1}{\sin\theta\left(\frac{\sin^2\psi}{I_A}+\frac{\cos^2\psi}{I_B}\right)}(p_{\phi}-\cos\theta p_{\psi})\right)^2 + \frac{1}{2k_BTI_C}p_{\psi}^2.$$

The integrals in the classical rotational partition function $q_{rot}$ can then be evaluated using again with $\int_{-\infty}^{\infty} e^{-ax^2} dx = \sqrt{\pi/a}$, and $\int_0^{\pi} \sin\theta d\theta = 2$, $\int_0^{\pi} d\phi = 2$, $\int_0^{\pi} d\psi = 2$ to yield

$$q_{rot} = \frac{\pi^2}{\sigma} \sqrt{\frac{8\pi k_B T I_A}{h^2}} \sqrt{\frac{8\pi k_B T I_B}{h^2}} \sqrt{\frac{8\pi k_B T I_C}{h^2}}.$$